\documentclass[11pt,twoside]{article}

\usepackage{asp2006}
\usepackage{epsf}
\usepackage{psfig}
\usepackage{lscape}

\newcommand{\Ha}{H$\alpha$}
\newcommand{\Hb}{H$\beta$}
\newcommand{\Hi}{{\sc H$\,$i}}
\newcommand{\Hii}{{\sc H$\,$ii}}
\newcommand{\Oiii}{[{\sc O$\,$iii}]}
\newcommand{\Oii}{[{\sc O$\,$ii}]}

\newcommand{\Nii}{[{\sc N$\,$ii}]}

\newcommand{\sauron}{{\texttt {SAURON}}}
\newcommand{\oasis}{{\texttt {OASIS}}}
\newcommand{\galex}{{\it GALEX}}
\newcommand{\spitzer}{{\it Spitzer}}

\newcommand{\eg}{e.g.,}

\begin{document}
\title{Recent Star Formation in Nearby Early-type Galaxies}   

\author{Marc Sarzi,\altaffilmark{1} Roland Bacon,\altaffilmark{2}
       Michele Cappellari,\altaffilmark{3}
       Roger~L. Davies,\altaffilmark{3}\\ 
       P.~Tim de Zeeuw,\altaffilmark{4} Eric Emsellem,\altaffilmark{2}
       Jes\'us Falc\'on-Barroso,\altaffilmark{5}\\ 
       Davor Krajnovi\'{c},\altaffilmark{3} Harald
       Kuntschner,\altaffilmark{6}
       Richard~M. McDermid,\altaffilmark{4}
       Reynier~F. Peletier,\altaffilmark{7} and Glenn van de
       Ven\altaffilmark{8}}

\affil{$^{\rm1}$Centre for Astrophysics Research, University of
Hertfordshire, Hatfield, AL10~9AB, UK}
\affil{$^{\rm2}$Universit\'e de Lyon 1, CRAL, Observatoire de Lyon, 9
av. Charles Andr\'e, 69230 Saint-Genis Laval, France}
\affil{$^{\rm3}$Sub-Department of Astrophysics, University of Oxford,
Denys Wilkinson Building, Keble Road, Oxford OX1~3RH, UK}
\affil{$^{\rm4}$Sterrewacht Leiden, Universiteit Leiden, Postbus
9513, 2300~RA, Leiden, The Netherlands}
\affil{$^{\rm5}$European Space and Technology Centre, Keplerlaan 1,
2200~AG Noordwijk, The Netherlands}
\affil{$^{\rm6}$Space Telescope European Coordinating Facility,
European Southern Observatory, Karl-Schwarzschild-Str.~2, 85748
Garching, Germany}
\affil{$^{\rm7}$Kapteyn Astronomical Institute, University of
Groningen, NL-9700 AV Groningen, The Netherlands}
\affil{$^{\rm8}$Institute for Advanced Study, Einstein Drive,
Princeton, NJ~08540, USA}

\begin{abstract} 
Motivated by recent progress in the study of early-type galaxies owing
to technological advances, the launch of new space telescopes and
large ground-based surveys, we attempt a short review of our current
understanding of the recent star-formation activity in such intriguing
galactic systems.
\end{abstract}


\section{Introduction}

Early-type galaxies represent one of the biggest challenges to
our understanding of galaxy formation and evolution, in particular as
far as their star-formation history is concerned. Studies of the
nearby galaxy population can provide useful clues to solve this
problem, by setting tight constraints on the level of their most
recent star-formation activity.
Traditionally, early-type galaxies used to be regarded as simple and
old stellar systems mostly devoid of gas or dust. By the turn of the
century, however, evidence was already mounting that a substantial
fraction elliptical and lenticular galaxies do show conspicuous gas
and dust reservoirs \citep[\eg][and references therein]{Gou94,Mac96},
and the presence of younger stellar components \citep[\eg][and
references therein]{Tra00}.
In the last years tremendous progress has been made in uncovering the
recent star-formation history of early-type galaxies, owing in
particular to the advent of new technologies such as integral-field
spectroscopic units or more sensitive receivers of radio and mm
frequencies, the launch of space observatories such as \galex\ and
\spitzer\ allowing to probe wavelength domains otherwise inaccessible
from the ground, and to large spectroscopic campaigns such as the
Sloan Digital Sky Survey (SDSS).
In this paper we attempt to review such recent efforts, without any
pretence of completeness. We will first focus on detailed studies of
the stellar populations of early-type galaxies at different
wavelengths, in particular with the \sauron\ integral-field unit in
the optical and with the \galex\ and \spitzer\ in the UV and Mid-IR
domain (\S\ref{sec:stars}). We will then turn to the reservoirs of
neutral, molecular and ionised gas in nearby elliptical and lenticular
galaxies, and consider evidence for on-going star formation in them
(\S\ref{sec:gas}). Finally we will consider some results based on the
analysis of large sample of early-type galaxies with SDSS data
(\S\ref{sec:largervolumes}), and conclude by discussing how all these
findings could connect with each other (\S\ref{sec:conclusions}).

\section{Stellar Population Studies and Evidence of Recent Star Formation}
\label{sec:stars}

\subsection{Optical - SAURON Results}
\label{subsec:Optical_stars}

Many studies have shown that early-type galaxies are not exclusively
formed by old stars \citep[e.g.,][and references therein]{Tho05}.
The \sauron\ survey of early-type galaxies \citep{deZ02} has further
allowed to find the location of different stellar sub-populations, in
particular of young stars, and to link such structures to the
photometry and kinematics of early-type galaxies.
Using maps for the strength of the \Hb, Mg$b$ and Fe5015 absorption
lines in a representative sample of 48 early-type galaxies in
Kuntschner et al. (in preparation) we reinforce previous indications
that the integrated stellar populations of most elliptical galaxies
display old stellar ages whereas lenticulars exhibit a wider spread of
ages, and likewise for galaxies in clusters compared to objects in a
lower density environment.
Very young stars, less than $\sim2$-Gyr-old, are found to dominate the
entire \sauron\ field-of-view of view in 4 objects in the \sauron\
sample, whereas in the remaining galaxies the young populations appear
to be either distributed in extended disks or to cluster towards the
centre. In fact, such young nuclear populations are also likely to
have formed in small stellar disks, since the presence of young stars
in the centre correlates very well with that of small (less than
$\sim$ 500pc) stellar components that are kinematically distinct from
the rest of the galaxy \citep{McD06}.

The connection between stellar ages and kinematics extends also beyond
the central regions.
In \citet{Ems07} we assess the overall level of rotational support in a
galaxy adopting a quantity, $\lambda_{\rm R}$, that is closely related
to the specific angular momentum of a galaxy.
In the \sauron\ sample, galaxies with $\lambda_{\rm R} \le 0.1$ form a
distinct class characterised by little or no global rotation and the
presence of kpc-scale kinematically decoupled cores. Opposed to such
slowly rotating objects are galaxies that either display faster global
rotation or that are consistent with being systems supported by
rotation that are viewed at small inclinations (see also the
contribution by Falc\'on-Barroso for an illustration of these two
kinds of objects).
All \sauron\ galaxies that formed stars recently (which amount to 25\%
of the sample and include either objects with luminosity-weighted mean
ages below 6.5 Gyr or with young disk and nuclei), are fast rotators,
whereas slow rotators do not show evidence of secondary star-formation
events.
In fact, beside the presence of young disks and nuclei, fast rotators
can also display extended, Mg$b$-enhanced structures that appear to be
flatter than the stellar isophotes \citep{Kun06}. Although not
obviously younger than the rest of the galaxy, such stellar
sub-components most likely originated in a secondary star-formation
event, consistent with the finding that fast rotators host dynamically
distinct stellar components characterised by large angular momentum
and radial anisotropy, as observed in stellar disks \citep{Cap07}.

\subsection{UV - GALEX Results}
\label{subsec:UV_stars}

Although the strength of optical absorption lines such as those probed
by \sauron\ can provide sensible constraints on the presence of
younger stars when compared to the predictions of {\it single\/}
stellar population models, deriving the mass and age of young stellar
components from line-strength indices alone is an exercise that is
freighted with degeneracies.
In this respect probing the UV light allows to weight and date even a
relatively small fraction of very young stars, since the UV output of
a stellar population varies dramatically over the first few 100 Myr of
its history.
In particular, whereas the far-UV (between 1350--1750\AA) light
is also sensitive to the radiation from old hot helium-burning
horizontal branch stars, the near-UV radiation (NUV, between
1750--2750\AA) traces particularly well the presence of young stars.

The potential of such UV information to constrain the age and fraction
of young stars of early-type galaxies has been recently shown by
\citet{Joe07}, who combined UV observations from \galex\ with
ground-based optical imaging for the E4 galaxy NGC~2974.  They find
blue NUV-optical colour indicative of the presence of very young
stars, less than 500-Myr-old, both near the centre and in an outer
ring. The star-formation rate and mass fraction of young stars peaks
in the outer ring, reaching ${\rm 2 \times 10^{-3} M_{\odot} yr^{-1}
kpc^{-2}}$ and 0.1\% respectively, whereas overall 1\% of the stars in
NGC2974 formed in the last 500 Myr.
Using the mass model of \citet{Kra05}, \citeauthor{Joe07} also show
that the location of the outer UV ring, as well as the central rings
traced by the \Oiii$\lambda\lambda4959,5007$ emission in the \sauron\
data for this galaxy, is consistent with the position of orbital
resonances that would be induced by the presence of a rotating
large-scale bar, which is driving the observed star formation in these
regions.
As most of the early-type galaxies in the \sauron\ sample are being
observed with \galex, it will be interesting to see how many more
objects will exhibit evidence of recent star formation in the UV
light, as in the case of NGC~2974.

\subsection{MidIR - Spitzer Results}
\label{subsec:midIR_stars}

At the opposite end of the optical spectrum, mid-IR (MIR) observations
provide an alternative way to investigate the stellar populations of
early-type galaxies. The MIR spectra of passively evolving stellar
populations are affected by the presence of mass-losing giant stars,
in particular AGB stars. The dusty envelopes of these stars reradiate
the heat received from their parent stars in a characteristic broad
region between 9 and 12 $\mu$m, mostly due to silicate grains.
As a stellar population ages, the photospheric MIR continuum fades
less quickly than the circumstellar 10 $\mu$m emission, so that the
relative strength of the silicate features can be used to trace the
age of the population.

The advent of \spitzer\ has recently allowed to routinely detect the
10 $\mu$m circumstellar emission in early-type galaxies, leading for
instance \citet{Bres06} to confirm that 82\% of their Virgo cluster
galaxies have been passively evolving since their
formation. \citet{Breg06} even finds that early-type galaxies
generally show old MIR-inferred ages even when optical analysis based
on absorption line strengths indicates relatively young
luminosity-weighted mean ages.
Although these results seem at odds with the larger fraction of
\sauron\ galaxies with evidence of recent star formation, we need to
keep in mind that both these MIR studies concentrated only on the
brightest end of the local early-type galaxy population, and that the
MIR age estimates are still subject to a severe age-metallicity
degeneracy \citep{Bres06}.

\section{Gaseous Reservoirs and Evidence of On-going Star Formation}
\label{sec:gas}

If stellar population studies provides evidence of recent star
formation in nearby early-type galaxies, we also ought to find gas
reservoirs capable of fuelling such activity before and while star
formation occurs.
Furthermore, the detection of molecular gas and of emission from
\Hii-regions should correspond to the presence of the youngest of
stars.

\subsection{Neutral Gas}
\label{subsec:neutral_gas}

A number of shallow \Hi\ surveys have shown already that early-type
galaxies can have massive (up to ${\rm 10^{10} M_\odot}$) and very
extended (up to 200 kpc in size) disks of neutral hydrogen
\citep[e.g.,][and references therein]{Oos02}. Such investigations
could only detect the most \Hi-rich galaxies, however, which prompted
\citet{Mor06} to set out to explore the complete \Hi-mass distribution
of early-type galaxies, starting from the representative \sauron\
sample.
So far, neutral gas has been found in 70\% of the 12 {\it field\/}
galaxies initially observed by \citet{Mor06}, with masses between few
times ${\rm 10^{6} M_\odot}$ and just over ${\rm 10^{9} M_\odot}$.
The neutral material in these objects seems also connected to the
ionised gas observed by \sauron\ in the optical regions, since all
galaxies where \Hi\ is detected also contain ionised gas, whereas no
\Hi\ is found in galaxies without ionised gas. Additionally, in the
most gas-rich systems the neutral and ionised material display a
similar kinematics. Considering that the presence of ionised-gas may
not always lead to star formation (see \S\ref{subsec:ionised_gas}),
these early results suggest that sufficient material to explain the
observed amount of recent star formation could be generally present
around early-type galaxies, and that gas accretion does not happen
exclusively in peculiar early-type galaxies.

\subsection{Molecular Gas}
\label{subsec:molecular_gas}

In order to form stars gas ultimately has to cool and condense,
leading also to the formation around dust grains of molecules such as
carbon monoxide.
Similarly to the case of \Hi, a number of CO emission surveys
\citep[see, e.g., the compilation by][]{Bet03} have revealed that
early-type galaxies can show substantial amounts of CO emission, which
generally traces regularly rotating molecular disks
\citep[e.g.,][]{You02,You05}.
More recently, \citet*{Com07} have followed up with single-dish
observations 43 out of the 48 early-type galaxies in the \sauron\
sample, detecting CO emission in 28\% of them.
This fraction is somewhat smaller than previously reported, but this
is because \citeauthor{Com07}, contrary to earlier studies, surveyed
also bright early-type galaxies that have a relatively lower molecular
gas content.
All galaxies with CO emission show evidence of recent star formation
in their central regions from \sauron\ or \oasis\ data, and have radio
continuum to far-IR (FIR) flux ratios that further supports the case
for on-going star formation.
In fact, the CO-rich \sauron\ early-type galaxies appear to extend the
well-known relation between the star-formation rate (SFR) surface
density and the surface density of ${\rm H_2}$ by two orders of
magnitude down in SFR.

In addition to mm-observations, the MIR domain provides additional
ways to trace even tiny amount of star-formation activity, in
particular through the detection of emission from polycyclic aromatic
hydrocarbon (PAH) features \citep[e.g.,][]{Kan05}. Although most
investigations with \spitzer\ have focussed on the brightest end of
the early-type population, the first results are nonetheless
promising. For instance, \citet{Bres06} finds evidence of recent star
formation in 12\% of their Virgo cluster sample, whereas \citet{Pan07}
succeed in quantifing the epoch ($\sim 200$ Myrs ago) and intensity
(adding in average ${\rm \sim 1 M_{\odot} yr^{-1}}$) of the recent
star-formation event in the central regions of the SB0 NGC~4435,
adding just about 1.5\% of the stellar mass in these regions.

\subsection{Ionised Gas}
\label{subsec:ionised_gas}

Optical nebular emission has traditionally been used to trace the warm
($\sim10^4$K) component of the interstellar medium (ISM) and identify
star formation in \Hii\ regions. Ionised-gas emission in early-type
galaxies is generally quite weak, however, so that its detection
requires a careful subtraction of the stellar continuum.

By introducing a novel technique to address this issue, in
\citet{Sar06} we could reach down to a detection limit of just 0.1\AA\
for the equivalent width (EW) of the emission lines and detect
extended ionised-gas emission in 75\% of the early-type galaxies in
the \sauron\ representative sample, confirming that these systems
contain gas more often than not.
Although the limited \sauron\ wavelength range does not allow an
extensive line-diagnostic analysis, very low \Oiii/\Hb\ line ratios
nonetheless suggest on-going star formation in at least 4 objects, or
8\% of the sample. In fact, that stars are forming in these galaxies
is proven also by the finding of disk-like gas kinematics, regular
dust morphology, young stellar populations and molecular gas in all of
them.
Interestingly, very young stars, CO emission and even PAH features
(Shapiro, private communication) can be found in objects where the
warm gas displays very high \Oiii/\Hb\ ratios \citep[see, e.g.,
NGC3489 and NGC3156 in][]{Sar06}. This finding suggests that as
ionised gas in early-type galaxies is also subject to different
sources of excitations than O-stars, it may not always be a good
tracer of star-formation, in particular if this occurs at a low
level. On the other hand, all the \sauron\ early-type galaxies with
compelling evidence of on-going star formation share a relaxed gas
kinematics, displaying in particular small gas velocity dispersions.

This common kinematic characteristic is found exclusively in fast
rotators in the \sauron\ sample, however, suggesting that not all
early-type galaxies are equally likely to form stars. In particular,
the spatial correlation with the hot ($\sim10^7$K) X-ray emitting
phase of the ISM \citep[][and references therein]{Sar07} suggests that
the fate of ionised gas in the most massive and slowly-rotating
early-type galaxies is to evaporate in the hot medium rather than to
condense and form stars \citep[see also][for a quantitative
analysis]{Nip07}. Preventing star formation in the most massive
objects consistently over time through the interaction with the X-ray
gas could explain the low frequency of CO and PAH detection in these
systems (\S\ref{subsec:molecular_gas}), and is in agreement with the
notion that the most massive galaxies host also the oldest stellar
populations.

\section{Results from Larger Volumes}
\label{sec:largervolumes}

If the previous efforts show with tremendous detail that a substantial
fraction of nearby early-type galaxies is or has been recently forming
stars, such local studies can only offer a sketchy picture of the
recent star-formation history of early-type galaxies in the local
Universe. 
Large surveys such as the SDSS, on the other hand, allow to
investigate of much larger number of early-type galaxies, although by
generally offering only the integrated photometric and spectroscopic
properties of these objects.

Some of the most dramatic results in this subject have been obtained
by combining the SDSS optical photometry with NUV images obtained
with \galex. Out of 39 bright early-type galaxies from the sample of
\citet{Ber03}, \citet{Yi05} found 6 objects with NUV -- r-band colours
consistent with the formation in the last Gyr of stars amounting to
approximately 1-2\% of the stellar mass. \citet{Kav06} considerably
extended \citeauthor{Yi05} first exercise to analyse a sample
$\sim2100$ morphologically-selected early-type galaxies from the SDSS
with ${\rm z < 0.11}$, finding that at least $\sim30\%$ of their
objects had formed between 1 to 3\% of their mass in the last Gyr,
mostly between 300 and 500 Myrs ago.
In an even larger effort, \citet{Sch07} have compiled a catalogue of
$\sim16000$ morphologically-selected early-type galaxies with redshift
between ${\rm 0.05 < z < 0.10}$ and SDSS, \galex\ and 2MASS
photometric and spectroscopic data, finding not only similar fractions
of galaxies with recent star formation, but also that $\sim 4\%$ of
their early-type galaxies are currently forming stars based on their
emission-line ratios.

The recent work of \citet{Gra07} further complements these
studies. \citeauthor{Yi05} and \citeauthor{Kav06} concentrated on
red-sequence objects without strong emission to avoid AGN
contamination in the UV light, whereas \citeauthor{Sch07} include
galaxies with a broader range of optical colours and with gas emission
but conservatively deem as quiescent objects without reliable \Ha,
\Hb, \Nii$\lambda\lambda6548,6583$ and \Oiii\
emission. \citeauthor{Gra07} investigates more closely the
red-sequence early-type galaxy population with gas emission, by
including objects with relatively fainter gas emission as long as they
display \Ha\ and \Oii$\lambda\lambda3726,3729$ lines that indicates
LINER-like emission\footnote{That is, characterised by low-ionisation
emission, as in the case of the Low Ionisation Nuclear Emission
Regions first identified by \citet{Hec80}}, or alternatively, which
firmly exclude on-going star formation.
With such a sensitive distinction between galaxies with or without
ionised-gas \citet{Gra07} find that red-sequence galaxies with
LINER-like emission show integrated stellar populations that are
invariably younger than their quiescent counterparts, except at the
brightest and more massive end of the red-sequence where galaxies show
no sign of recent star formation.

\section{Conclusions}
\label{sec:conclusions}

How do all these finding combine together? Is it possible to draw a
consistent picture of the recent star-formation history of local
early-type galaxies based both on the detailed study of the closest
galaxies and on the analysis of integrated properties for much larger
samples?  Unfortunately, the answer is not quite yet.

Starting with the fraction of early-type galaxies that are presently
forming stars, at first glance the fact that 8\% of the \sauron\
sample display \Oiii/\Hb\ line ratios consistent with emission from
\Hii\ regions would seem in reasonable agreement with the finding that
4\% of the objects in the SDSS sample of \citeauthor{Sch07} display
the same behaviour. 
We have to keep in mind, however, that the \sauron\ sample is a
representative, but {\it incomplete\/} sample of the nearby early-type
population, in which fainter galaxies are particularly
underrepresented. As star formation could be more common in these
systems than in more massive galaxies (\S\ref{subsec:ionised_gas}),
the true fraction of early-type galaxies with emission associated to
\Hii\ regions could be much higher than presently found by \sauron.
Similarly, also the fraction of star-forming early-type galaxies
estimated from more complete SDSS samples should be regarded as a
lower limit, since the SDSS spectra can only detect the most intense
of starbursts. 
For instance, considering an EW of 0.8\AA\ for the faintest lines
detected in the SDSS data, only 1 of the 48 \sauron\ galaxies has
sufficiently strong emission within $1{\rm R_e}$ (corresponding to the
typical physical area subtended by the SDSS fibers) to be detected as
one of the star-forming early-type galaxies of \citeauthor{Sch07}.

As regards the presence of younger stellar components in early-type
galaxies, it is reassuring that \citet{Sch07} and \citet{Kav06}
estimate fractions of galaxies that are presently forming stars ($\sim
4\%$) and that contain populations younger than a few 100 Myr ($\sim
30\%$), respectively, which are consistent with each other given the
different timescales traced by the nebular and NUV emission.
On ther other hand, it is less clear how well the mass fractions and
ages of the young stellar components derived by \galex\ and SDSS data
agree with the \sauron\ observations.
Consider the fact that just 1\% of 500-Myr-old stars embedded in a
10-Gyr-old populations induces an increase of $\sim1$\AA\ in the
strength of the \Hb\ absorption line. 
Only 10\% of objects in the \sauron\ sample have a global (within
$1{\rm R_e}$) \Hb\ line-strength that exceeds by 1\AA\ what observed
in the oldest objects.
Although the discrepancy with the 30\% fraction reported by
\citeauthor{Kav06} could be reconciled considering that more \sauron\
objects show evidence of recent star formation (up tp 25\%,
\S\ref{subsec:Optical_stars}), until the mass and age fraction of
these young subcomponents is determined for a sufficiently complete
sample of nearby galaxies it will remain hard to ascertain whether
such \sauron\ objects correspond to the galaxies that experienced recent star
formation according to analyses based on \galex\ and SDSS data.

Despite these uncertainties, it is clear that with the exception of
the most massive objects most early-type galaxies have undergone some
degree of recent star formation. The gas reservoirs fuelling such
activity can now be detected with relative ease, and new techniques
have been developed to analyse data from different passbands to better
constrain the star formation history of galaxies.
The biggest challenge ahead is still to quantify more precisely the
extent of such star-formation activity and its impact in shaping the
observed stuctural and dynamical properties of these systems.
In fact, it is precisely to address these and other questions that we
have recently started a complete integral-field survey of the nearby
early-type galaxy population, nicknamed ATLAS$^{\rm 3D}$, which will
be complemented by single-dish mm-band observations and by deep
multi-wavelength optical imaging\footnote{Details can be found at {\tt
http://www-astro.physics.ox.ac.uk/atlas3d/\/}}.

\acknowledgements
Marc Sarzi is truly indebted to Johan Knapen for the opportunity to
present this small review at such an interesting conference, and
wishes to thank Lisa Young, Martin Bureau, Alessandro Bressan, Kristen
Shapiro, Sugata Kaviraj, Kevin Schawinski and Genevieve Graves for
many useful discussions.

\end{document}